\documentclass[aps,prl,superscriptaddress]{revtex4}
\usepackage{graphicx}
\usepackage{color}
\usepackage{amsmath,amssymb}
\usepackage{ulem}
\usepackage[caption=false]{subfig} % caption=false needs to be added for revtex4 for some reason
\usepackage{siunitx}
\usepackage{setspace}

\begin{document}

\newcommand{\avg}[1]{\langle #1 \rangle}
\newcommand{\bra}[1]{\langle #1 |}
\newcommand{\ket}[1]{| #1 \rangle}
\newcommand{\braket}[2]{\langle #1 | #2 \rangle}
\newcommand{\ketbra}[2]{| #1 \rangle\!\langle #2 |}
\newcommand{\up}{\uparrow}
\newcommand{\down}{\downarrow}
\newcommand{\eff}[1]{#1_{\rm eff}}
\renewcommand{\Re}{\mbox{Re}}
\renewcommand{\Im}{\mbox{Im}}

% Use the \preprint command to place your local institutional report
% number in the upper righthand corner of the title page in preprint mode.
% Multiple \preprint commands are allowed.
% Use the 'preprintnumbers' class option to override journal defaults
% to display numbers if necessary
%\preprint{}

%Title of paper
\title{Observing Coherence Effects in an Overdamped Quantum System}

\author{Y.-H. Lien}
\affiliation{Centre for Cold Matter, Imperial College, Prince Consort Road, London SW7 2BW, United Kingdom}
\affiliation{School of Physics and Astronomy, University of Birmingham, Edgbaston, Birmingham B15 2TT, United Kingdom}
\author{G. Barontini}
\email{g.barontini@bham.ac.uk}
\affiliation{Centre for Cold Matter, Imperial College, Prince Consort Road, London SW7 2BW, United Kingdom}
\affiliation{School of Physics and Astronomy, University of Birmingham, Edgbaston, Birmingham B15 2TT, United Kingdom}
\author{M. Scheucher}
\affiliation{Centre for Cold Matter, Imperial College, Prince Consort Road, London SW7 2BW, United Kingdom}
\affiliation{Vienna Center for Quantum Science and Technology, Atominstitut, TU Wien, 1020 Vienna, Austria}
\author{M. Mergenthaler}
\affiliation{Centre for Cold Matter, Imperial College, Prince Consort Road, London SW7 2BW, United Kingdom}
\affiliation{Department of Materials, University of Oxford, Parks Road, Oxford OX1 3PH, United Kingdom}
\author{J. Goldwin}
\email{j.m.goldwin@bham.ac.uk}
\affiliation{School of Physics and Astronomy, University of Birmingham, Edgbaston, Birmingham B15 2TT, United Kingdom}
\author{E. A. Hinds}
\email{ed.hinds@imperial.ac.uk}
\affiliation{Centre for Cold Matter, Imperial College, Prince Consort Road, London SW7 2BW, United Kingdom}

%\email[]{Your e-mail address}
%\homepage[]{Your web page}
%\thanks{}
%\altaffiliation{}
%\date{\today}

\begin{abstract}
It is usually considered that the spectrum of an optical cavity coupled to an atomic medium does not exhibit a normal-mode splitting unless the system satisfies the strong coupling condition, meaning the Rabi frequency of the coherent coupling exceeds the decay rates of atom and cavity excitations. Here we show that this need not be the case, but depends on the way in which the coupled system is probed. Measurements of the reflection of a probe laser from the input mirror of an overdamped cavity reveal an avoided crossing in the spectrum which is not observed when driving the atoms directly and measuring the Purcell-enhanced cavity emission. We understand these observations by noting a formal correspondence with electromagnetically-induced transparency of a three-level atom in free space, where our cavity acts as the absorbing medium and the coupled atoms play the role of the control field.

\end{abstract}

\maketitle

The oscillatory dynamics intrinsic to quantum systems can be harnessed in a number of ways for advanced applications in metrology and sensing. Coherent oscillations among quantum states are at the heart of existing time-frequency standards \cite{Ram90,Lud15}, atom interferometers \cite{Cro09,Hor12}, and quantum information processors (see, for example, Ref.~\cite{coherence}). Coherence in these systems is ultimately limited by dephasing which arises from the coupling of the isolated quantum system to the vast reservoir of states representing the environment. In some cases the system is open by design, for example to allow communication of information into and out of the system; in other cases it is unavoidable, due to losses or the influence of uncontrolled fluctuations in the reservoir. In either case the question naturally arises --- to what extent can the effects of quantum coherence survive in the presence of damping and decay?

Here we study a system comprising cold atoms coupled to a high finesse optical microcavity. The relatively small cavity mode volume results in a large electric dipole coupling between a single photon and an atom. This comes at the price of a fast optical decoherence rate due to transmission and losses from the cavity mode into external modes. Despite this, we show how the Hamiltonian eigenfrequencies of the system can be revealed through spectroscopic measurements. By probing the cavity with an incident field, we realise an analogy to electromagnetically induced transparency (EIT) in which the coherent atom-cavity coupling plays the role of the control field. The effective three-level system is formed by the lowest triplet of dressed states of the coupled atom-cavity complex. We refer to this as dressing-induced transparency (DIT), and it is complementary to the cavity-induced transparency first proposed in \cite{Rice96}, with the roles of cavity and atom here reversed. As described most systematically in \cite{Souza15}, a wide variety of physical systems exhibit formally similar dynamics, where destructive interference inhibits excitation under weak coupling (see also Table 1 of \cite{Pen14}). A quantum dot in a photonic crystal cavity has been used for optical switching between a pair of coupled waveguides \cite{Waks06,Faraon08}, giant optical nonlinearities have been predicted \cite{Auffeves07}, and a recipe for entangling atom pairs has been described \cite{Rossatto13}.

Our experiment differs from most realisations through the clear separation of the rates of coherent and incoherent processes --- there are three orders of magnitude between the atom and cavity decay rates, with the Rabi frequency near the geometric mean. This leads to a transparency line width which is deeply sub-natural, in the sense of being far below the bare cavity absorption line width. On top of this we exploit a feature of the dark state which so far appears to have gone unnoticed --- the full spectral response reveals the normal mode frequencies of the undamped system.  Because of this, we are able to go beyond previous work to observe an avoided crossing in the dressed cavity spectrum, which is usually taken as evidence of strong coupling, despite the complete overdamping of Rabi oscillations in our experiment. We also control the width of the DIT window through the density dependence of the atomic dipole. The analogy to EIT also highlights the existence of steep dispersion present in the system \cite{Shimizu02}. In contrast with conventional EIT of three-level atoms in optical cavities \cite{Wang00,Hernandez07,Mucke10,Kampschulte10,Albert11,Tanji11}, it is possible in our system to swap the roles of longitudinal and transverse dephasing rates, corresponding to decay of populations and coherences, respectively. When the probe laser is incident on the atoms directly (from the side of the cavity), DIT per se is not present. However, the spectral width of the narrow feature in this case is the same as for the dressing-induced transparency window with the same parameters, giving insights on the well-known Purcell effect describing coherent scattering into the cavity mode \cite{Pur46}. 

In the following we study both theoretically and experimentally the reflection spectra of our system when the cavity mode is probed and the fluorescence spectra when the atoms are probed. In each case we measure the system response when the cavity-atom and laser-atom detunings are varied. We show that the reflection spectra exhibit an avoided crossing while the fluorescence spectra do not. However the measured resonance line widths are the same. Our results show that even in a highly overdamped regime, quantum interference can be used to reveal aspects of the coherent coupling.

\section{Results}

\subsection{Theoretical framework}

A single two-level atom coupled to a single mode of an ideal optical cavity is described in the dipole and rotating wave approximations by the Jaynes-Cummings Hamiltonian \cite{Jay63} ($\hbar=1$),
\begin{eqnarray}
\hat{H} &=& \omega_{ \rm c}\,\hat{a}^\dagger\hat{a} + \omega_{ \rm a}\,\hat{\sigma}^\dagger\hat{\sigma}  -g\,(\hat{\sigma}^\dagger\hat{a} + \hat{\sigma}\hat{a}^\dagger) \quad.
\label{eq:H}
\end{eqnarray}
Here the operator $\hat{a}$ annihilates a photon from the cavity field and $\hat{\sigma}=\ketbra{\!\down\,}{\,\up\!}$ is the Pauli isospin operator lowering the atom from the excited state $\ket{\!\up\,}$ to the ground state $\ket{\!\down\,}$, $\omega_{ \rm c}$ ($\omega_{ \rm a}$) is the angular frequency of the uncoupled cavity resonance (atomic transition), and the atom-field coupling constant, $g$, is proportional to the atomic dipole moment and inversely proportional to the square root of the cavity mode volume. The Hamiltonian (\ref{eq:H}) conserves the total excitation number $\hat{N}_{\rm exc}=\hat{a}^\dagger\hat{a}+\hat{\sigma}^\dagger\hat{\sigma}$, and the simultaneous eigenstates of $\hat{H}$ and $\hat{N}_{\rm exc}$ consist of a dark ground state, $\ket{\!\!\down,0}$ obeying $\hat{H}\ket{\!\!\down,0}=0=\hat{N}_{\rm exc}\ket{\!\!\down,0}$, and a ladder of doublets consisting of superpositions of $\ket{\!\!\up,n-1}$ and $\ket{\!\!\down,n}$ with $\avg{\hat{N}_{\rm exc}}=n=1,2,3$, etc.

\begin{figure}
\centering\includegraphics[width=8cm]{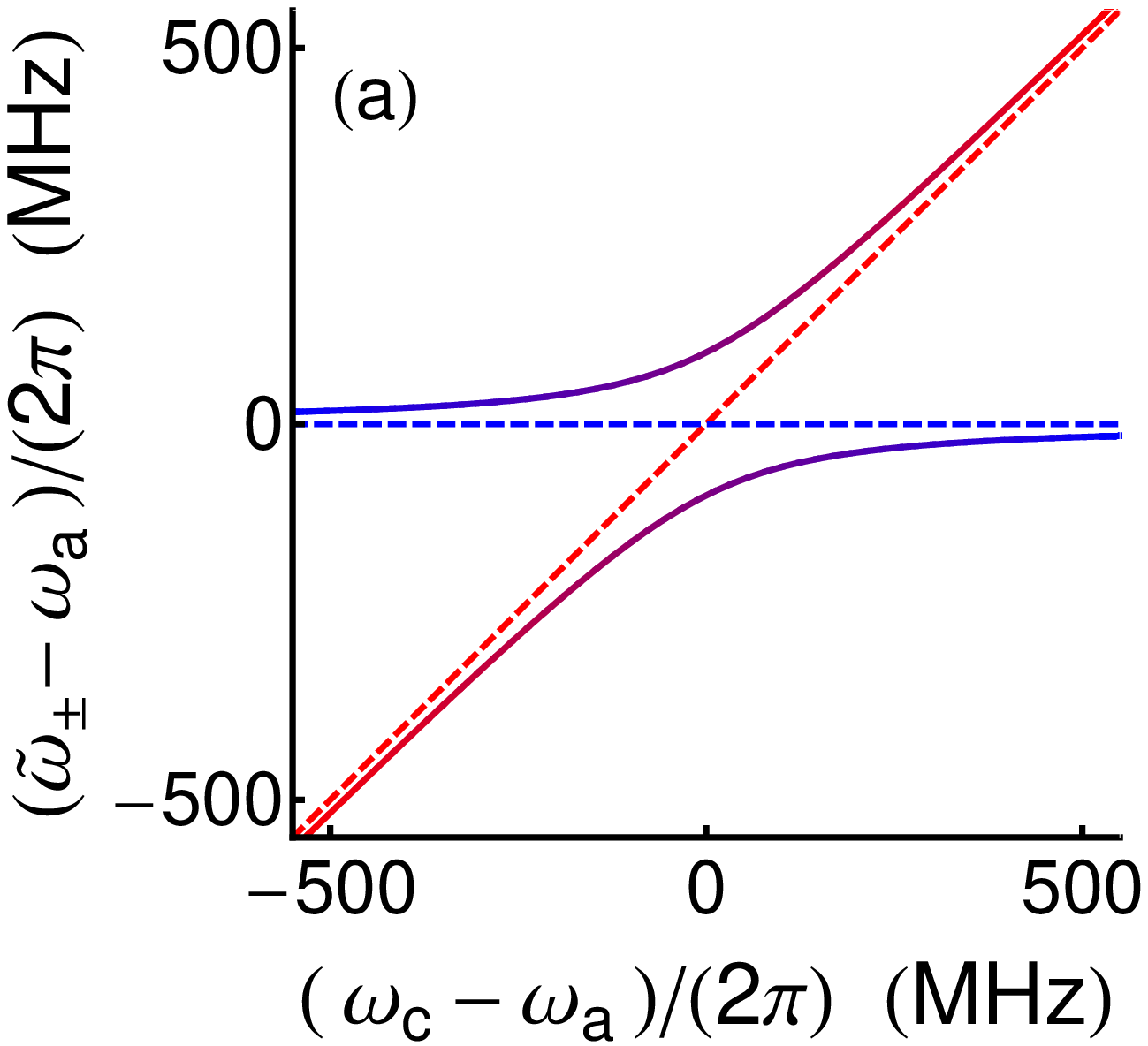}
\includegraphics[width=8cm]{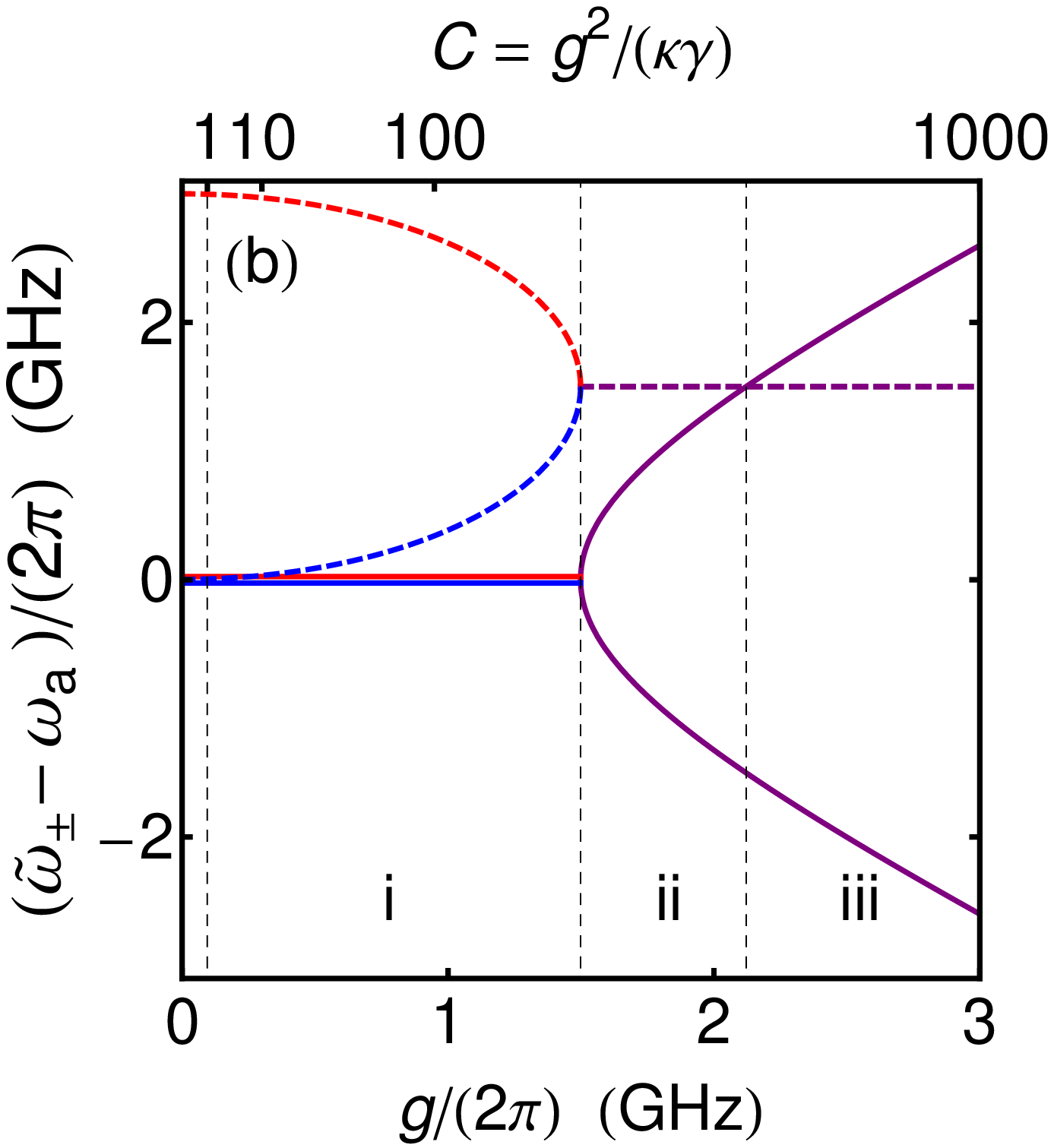}
\caption{{\bf Lowest-lying normal modes of the atom-cavity system.} (a) Eigenfrequencies $\omega_\pm$ of the undamped Hamiltonian $\hat{H}$. Dashed lines: uncoupled atom (blue), and cavity (red). Solid lines: avoided crossing of the eigenfrequencies $\omega_\pm$ of the dressed system, as given by Eq.~(\ref{eq:undampedeigens}) with $g/(2\pi)=95$~MHz. Colour indicates the relative amplitudes of the bare states $\ket{\!\!\up,0}$ (more blue) and $\ket{\!\!\down,1}$ (more red) in the eigenstate. (b) Complex eigenvalues $\tilde{\omega}_\pm$, given in Eq.~(\ref{eq:eigenfreqs}) for the non-Hermitian effective Hamiltonian $\hat{H}_{\rm eff}=\hat{H}-i\kappa\,\hat{a}^\dagger\hat{a}-i\gamma\,\hat{\sigma}^\dagger\hat{\sigma}$, taking $\omega_{ \rm c}=\omega_{ \rm a}$ and $\{\kappa,\gamma\}/(2\pi)\simeq\{3000,3\}\,$MHz. Solid curves: eigenfrequencies $\Re(\tilde{\omega}_\pm)$. Dashed curves: damping rates $-\Im(\tilde{\omega}_\pm)$. Zones (i), (ii) and (iii) correspond respectively to the Purcell regime, the intermediate regime and the strong coupling regime.}
\label{fig:eigenfreqs}
\end{figure}

Figure \ref{fig:eigenfreqs}(a) shows the eigenvalues for the lowest doublet ($n=1$) of the coupled system:
\begin{eqnarray}
\omega_\pm - \omega_{ \rm a} &=& \frac{\omega_{ \rm c}-\omega_{ \rm a}}{2} \pm \sqrt{g^2+\left(\frac{\omega_{ \rm c}-\omega_{ \rm a}}{2}\right)^2} \;.
\label{eq:undampedeigens}
\end{eqnarray}
The interaction lifts the degeneracy at $\omega_{ \rm c}=\omega_{ \rm a}$, splitting the two levels in the eigenspectrum by $2g$ at this point. In the time domain, this splitting corresponds to the Rabi oscillation between the states $\ket{\!\!\up,0}$ and $\ket{\!\!\down,1}$.

To describe real systems we need to account for decoherence processes affecting both the atom and cavity field. The probability for the bare atom to be in the excited state decays irreversibly through spontaneous emission at a rate $2\gamma$, and the mean number of photons in the bare cavity decays at a rate $2\kappa$ due to transmission and losses at the mirrors. The three rates $\{g,\kappa,\gamma\}$ quantify the strength of the light-matter interactions through the dimensionless cooperativity $C=g^2/(\kappa\gamma)$. The cooperativity represents the effective optical depth of the atom \cite{Hor03}, or the ratio of coherent scattering into the cavity mode to scattering into free space \cite{Pur46}. When $C\gtrsim 1$, single-atom detection is therefore possible either through the modified cavity spectrum or through laser-induced fluorescence driven from the side of the cavity \cite{Tho92, Hor03,Wilzbach06, Tep06, Tru07,Goldwin11}. In the experiment we discuss here, $\gamma/(2\pi)=3$~MHz, $\kappa/(2\pi)$ varies from $2.2$--$3.2$~GHz depending on mirror alignment (see Methods), and $g/(2\pi)$ is continuously adjustable from zero up to $345\,$MHz as explained below, so the cooperativity $C$ can be large even though the Rabi frequency is much smaller than one of the decay rates.

Intuitively, one might expect that the avoided crossing of Fig.~\ref{fig:eigenfreqs}(a) would not be resolved in an experiment with $\kappa\gg g$ because the underlying Rabi oscillation between $\ket{\!\!\up,0}$ and $\ket{\!\!\down,1}$ would be interrupted by fast, irreversible cavity decay from $\ket{\!\!\down,1}$ to $\ket{\!\!\down,0}$. To describe this quantitatively, we restrict ourselves to the Hilbert sub-space spanned by the three lowest uncoupled states, $\{\ket{\!\!\down,0},\ket{\!\!\down,1},\ket{\!\!\up,0}\}$, which is valid for weak excitation (in the sense $\avg{\hat{N}_{\rm exc}}\ll 1$), and we account for dissipation through an effective Hamiltonian \cite{Hegerfeldt93}, $\hat{H}_{\rm eff}=\hat{H}-i\kappa\,\hat{a}^\dagger\hat{a}-i\gamma\,\hat{\sigma}^\dagger\hat{\sigma}$. The eigenvalues, $\tilde{\omega}_\pm$, of this non-Hermitian Hamiltonian are complex:
\begin{eqnarray}\nonumber
\tilde{\omega}_\pm-\omega_{ \rm a} &=& \frac{(\omega_{ \rm c}-\omega_{ \rm a})-i(\kappa+\gamma)}{2} \\
&~&\quad \pm \sqrt{g^2+\left[\frac{(\omega_{ \rm c}-\omega_{ \rm a}) - i(\kappa-\gamma)}{2}\right]^2}\,.
\label{eq:eigenfreqs}
\end{eqnarray}
The real parts of $\tilde{\omega}_\pm$ give the resonance frequencies of the coupled atom-cavity system, while the imaginary parts give the corresponding line widths. These are plotted in Fig.~\ref{fig:eigenfreqs}(b) as a function of $g$ for the case of $\omega_{ \rm c}=\omega_{ \rm a}$. Note that $\Re(\tilde{\omega}_\pm)$ depends on the difference between the two uncoupled damping rates, so it is not generally the same as $\omega_\pm$ from Eq.~(\ref{eq:undampedeigens}).

In Fig.~\ref{fig:eigenfreqs}(b) we identify three distinct parameter regimes. In (i), $g>\sqrt{\kappa\gamma}$, i.e. the cooperativity $C>1$, but $g<(\kappa-\gamma)/2$. Here, the real parts of $\tilde{\omega}_\pm$ are unshifted at $\omega_{c}$ and $\omega_{a}$, and have widths of approximately $\kappa-g^2/\kappa$ and $\gamma\,(1+C)$. These eigenmodes are primarily photonic and atomic, respectively, as reflected by the colouring of the lines in Fig.~\ref{fig:eigenfreqs}(b). The appearance of $C$ in the line width of the atomic feature is a direct signature of the enhanced spontaneous emission rate of the atom, as first predicted by Purcell \cite{Pur46}. We therefore refer to this as the Purcell regime; this is the operating regime for all of the experiments described here. In region (ii), above $g=(\kappa-\gamma)/2$ where there is an exceptional point \cite{Kato66}, the eigenfrequencies separate and the damping rates merge to the mean value $(\kappa+\gamma)/2$. Some authors call this strong coupling, but the normal modes are not resolved until the splitting exceeds the width. This consideration gives rise to a second, more stringent definition of strong coupling, namely, (iii): $g^2>(\kappa^2+\gamma^2)/2$. In this regime, multiple Rabi oscillations occur on average before the excitation decays. This definition of strong coupling is reminiscent of Rayleigh's criterion defining the resolution limit of an imaging system under incoherent illumination, and is the definition which we adopt here. 

\subsection{Probing the dressed system}

We now show that, when a coherent field probes the coupled system, quantum coherence between the atom and cavity can lead to quantum interference and narrow spectral features even in regime (i), where our experiment lies. That is the main focus of this article.

\begin{figure}
\centering
\includegraphics[width=0.8\textwidth]{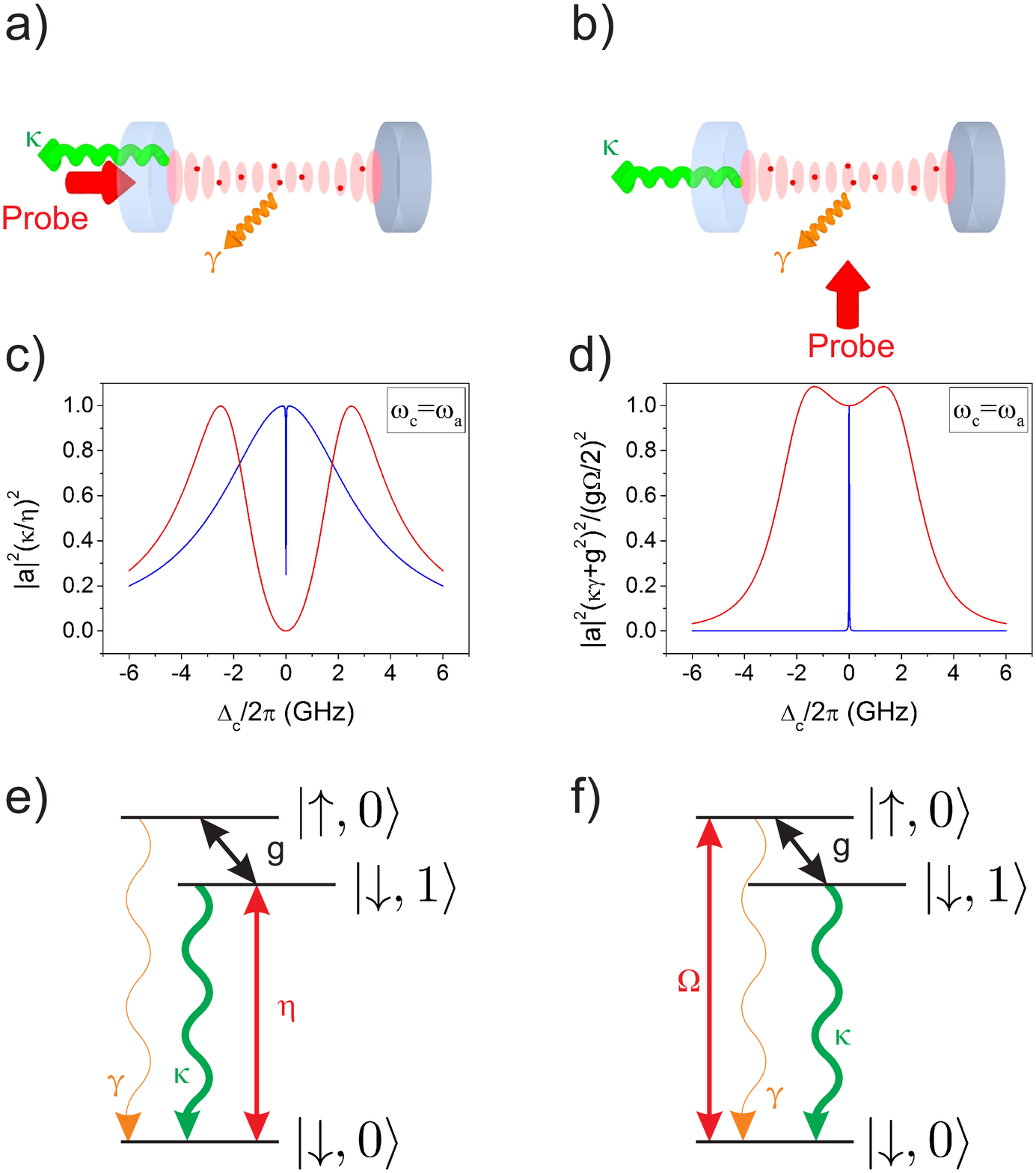}
\caption{{ \bf Schemes for probing the dressed atom-cavity system.} (a) Experimental configuration for driving the cavity mode. The field is probed through the cavity reflection. (b) Experimental configuration for driving the atoms. The field inside the cavity is probed by measuring light emitted through one of the cavity mirrors. (c) Number of intracavity photons, $|a|^2$, versus detuning for configuration (a). The narrow dip (blue) is for $C=1$ (i.e. $g/(2\pi)=95\,$MHz, $\kappa/(2\pi)=3\,$GHz,$\gamma/(2\pi)=3\,$MHz), while the wide one (red) is for strong coupling  -- $g/(2\pi)=2.5$\,GHz -- which lies in zone (iii) of Fig.~\ref{fig:eigenfreqs}(b). We scale the photon number by $(\kappa/\eta)^2$ (a very large number in our case) to obtain a curve that is independent of the probe strength $\eta$.  (d) Intracavity photon number (suitably scaled again) as a function of detuning for configuration (b). The narrow peak (blue) is for $C\sim 1$. The wide peak (red, $g/(2\pi)=2.5$\,GHz) is only just beginning to exhibit a dip at strong coupling. (e) and (f) Effective three-level systems corresponding to configurations (a) and (b) respectively. The cavity is assumed to be red-detuned from atomic resonance ($\omega_{ \rm c}<\omega_{ \rm a}$), leading to a ladder system in (e) and a $\Lambda$ system in (f); for blue detuning the configurations are reversed. For either detuning, the system supports a DIT coherence window in (e) and does not in (f).}\label{fig:analogy}
\end{figure}

In the laboratory, we excite the coupled atom-cavity system in two different ways. First, we probe the dressed cavity with coherent (laser) light through one of the mirrors, as illustrated in Fig.~\ref{fig:analogy}(a). This adds a driving term to $\hat{H}_{\rm eff}$ equal to $-i\eta\,(\hat{a}-\hat{a}^\dagger)$. Here $\eta^2=2\kappa_T j$, where $j$ is the number of photons per unit time driving the cavity mode and $\kappa_T$ is the contribution of $\kappa$ due to transmission through the input mirror \cite{Hor03}. In the interaction picture rotating at the probe frequency $\omega_{ \rm p}$, the equation of motion for the operator $\hat{\sigma}$ can be derived from the modified Heisenberg evolution equation, $d\hat{\sigma}/dt=i[\hat{H}_{\rm eff},\hat{\sigma}]$, and similarly for $\hat{a}$:
\begin{eqnarray}
\frac{d}{dt}\,\sigma &=& -(\gamma-i\Delta_{ \rm a})\,\sigma + iga \\
\frac{d}{dt}\,a &=& -(\kappa-i\Delta_{ \rm c})\,a + ig\sigma + \eta \quad,
\end{eqnarray}
where $a=\avg{\hat{a}}$ and $\sigma=\avg{\hat{\sigma}}$, and the detunings are defined by $\Delta_{\rm a,c}=\omega_{ \rm p}-\omega_{a,c}$. Here, we have assumed weak excitation of the atom, so that $\avg{\hat{\sigma}^\dagger\hat{\sigma}}\ll 1$. The steady state solution for the amplitude of the (coherent state) field in the cavity $a$ has been given in \cite{Souza15},
\begin{eqnarray}
a &=& \frac{\eta}{(\kappa-i\Delta_{ \rm c}) + g^2/(\gamma-i\Delta_{\rm a})} \quad.
\label{eq:ar}
\end{eqnarray}
So far, we have been considering a single atom with a fixed coupling $g$ to the cavity field, and therefore a fixed cooperativity. In the experiments however there is an ensemble of atoms simultaneously coupled to the cavity, and this results in a collective coupling $\eff{g}=g\sqrt{\eff{N}}$, where $\eff{N}$ is the volume integral of the atom number density, weighted by the intensity distribution of the cavity mode \cite{Car99, Goldwin11}. The expression for $a$ in Eq.~(\ref{eq:ar}) remains valid as long as we understand $g$ to mean $\eff{g}$.

Figure~\ref{fig:analogy}(c) shows the mean number of photons in the cavity as a function of laser-cavity detuning $\Delta_{ \rm c}$.  The photon number $|a|^2$ is scaled by $(\kappa/\eta)^2$ in order to obtain a curve that is independent of the probe strength $\eta$. The atom-cavity coupling is most pronounced when the atom and the cavity resonate simultaneously with the driving field, so we have set $\omega_{ \rm c}=\omega_{ \rm a}$ here. The wide double peak (red) is calculated for $g/(2\pi)=2.5\,$GHz, which lies in the region of strong coupling, indicated by (iii) in Fig.~\ref{fig:eigenfreqs}(b). This curve shows the two spectroscopically-resolved, coupled atom-cavity states. The curve with a narrow central dip (blue) has the coupling reduced to $g/(2\pi)=95\,$MHz ($C=1$), which places it at the left of zone (i) in Fig.~\ref{fig:eigenfreqs}(b). The two eigenvalues are not spectroscopically resolved in the sense of region (iii), but they are still apparent as a broad peak with a narrow hole, made visible by quantum coherence.

The spectra in Fig.~\ref{fig:analogy}(c) can also be understood in the language of EIT, as was previously noted for the complementary case where $\kappa\ll g<\gamma$ \cite{Rice96}. Interest in the transition from EIT to Autler-Townes splitting has recently grown \cite{Ani11,Pen14}, and in a cavity QED system this transition is manifest as a crossover from DIT to vacuum Rabi splitting. The analogy between DIT and conventional EIT is further illustrated in Fig.~\ref{fig:analogy}(e). Depending on the detuning between the cavity and atomic resonances, the three lowest dressed states of the cavity QED system form either ladder- or $\Lambda$-type configurations. In this analogy the cavity plays the role of the absorber, with the atomic dipole (proportional to $\sigma+\sigma^*$) acting as the coupling field which induces the transparency. The cavity field (proportional to $a+a^*$) drives the $\ket{\!\!\down,0}\leftrightarrow\ket{\!\!\down,1}$ transition, whose `natural' line width is set by the cavity damping rate, $\kappa$, and the vacuum-induced coupling on $\ket{\!\!\down,1}\leftrightarrow\ket{\!\!\up,0}$ is subject to dephasing at the rate $\gamma$, returning the system to the absolute ground state $\ket{\!\!\down,0}$. Whether the dressed system maps onto a ladder or $\Lambda$ configuration depends on the cavity-atom detuning and the method of probing. For our case, where $(g/\kappa)^2\ll 1$, direct cavity probing exhibits a narrow DIT window with sub-natural line width equal to $\gamma\,(1 + C)$, which is much narrower than the natural width $\kappa$ of the cavity being probed. In this expression we can identify $\gamma\,C$ as the coherent scattering rate predicted by Purcell and $\gamma$ as an effective dephasing rate. As long as the former exceeds the latter, meaning precisely that $C$ exceeds 1, a transparency window emerges in the cavity spectrum. Similarly, the depth of this window depends on $C$, and we are able to observe these effects experimentally because we can adjust the cooperativity in the regime where $C\ge1$ by varying the density of atoms in the cavity.

In a second method of probing the coupled system, we drive the atoms --- rather than the cavity --- by illuminating them with a near-resonant beam propagating transverse to the cavity axis, as illustrated in Fig.~\ref{fig:analogy}(b). This amounts to setting $\eta=0$ and adding a term $-(\Omega/2)(\hat{\sigma}+\hat{\sigma}^\dagger)$ to the Hamiltonian $\hat{H}_{\rm eff}$, with $\Omega$ being the probe Rabi frequency. The corresponding steady-state solution for the cavity field is then,
\begin{equation}\label{eq:af}
a=\frac{-g\,\Omega/2}{(\kappa-i\Delta_{ \rm c})(\gamma-i\Delta_{ \rm a})+g^2}\quad.
\end{equation}
Figure~\ref{fig:analogy}(d) shows the calculated number of photons in the cavity for this configuration as a function of the same laser frequency scan, again with $\omega_{ \rm c}=\omega_{ \rm a}$.  In this case we have scaled the photon number by $(g^2+\kappa\gamma)^2/(g\,\Omega/2)^2$ to obtain a curve that is independent of the probe strength $\Omega$. The narrow resonance at  $g/(2\pi)=95\,$MHz ($C=1$, blue) does not reveal that the system has two modes, and even with strong coupling at $g/(2\pi)=2.5\,$GHz (red), the normal modes are barely resolved, in contrast with the cavity-probed case. In the DIT language, the roles of atom and cavity are interchanged here, with the atoms acting as the absorber. Now the cavity field in Fig.~\ref{fig:analogy}(d) and (f) shows no DIT effect even when the condition $C>1$ is satisfied, because the dephasing rate is equal to $\kappa$ and the condition $\kappa\gg\gamma$ prevents the formation of a transparency window in the $\ket{\!\!\down,0}\leftrightarrow\ket{\!\!\up,0}$ transition \cite{Fleisch05}.

\subsection{Experimental observations}

In the experiment we measure the flux of photons travelling from the cavity to a single photon counting module (see Methods). This flux, $j_{\rm out}$, is related to the intracavity field amplitude, $a$:
\begin{eqnarray}\label{eq:flux}
j_{\rm out} &=& \left|-\sqrt{R_1\,j}+\sqrt{R_2\kappa}\,a\right|^2\quad,
\end{eqnarray}
where $R_1$ is the power reflection coefficient of the intput-output coupling mirror and $R_2$ is that of the back mirror. Here we have used the fact that $R_1\approx R_2\approx 1$, and that losses at mirror 1 are negligible compared to transmission in our experiment.

\begin{figure}
\centering\includegraphics[width=\textwidth]{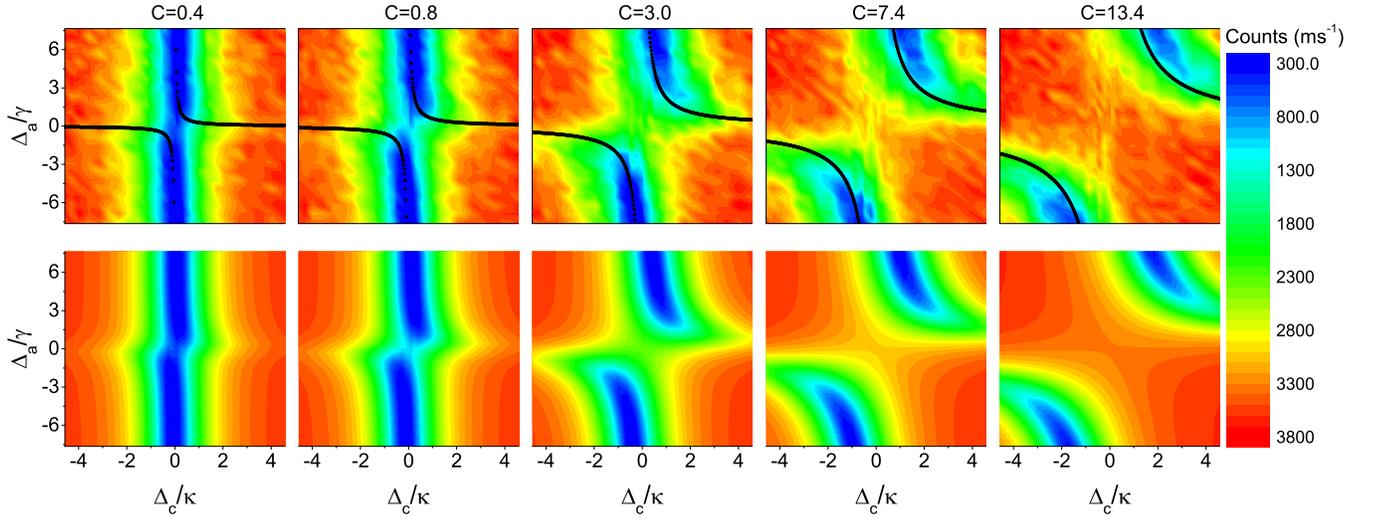}
\caption{{\bf Probing the system by driving the cavity mode.} Two dimensional scans of the cavity reflection signal in the configuration of Fig.~\ref{fig:analogy}(a) as a function of the detuning from the cavity ($\Delta_{ \rm c}$) and from the atomic ($\Delta_{ \rm a}$) resonances for different values of the cooperativity $C$. Upper row: Experimental spectra. Note that the axes are normalised to the respective damping rates. The colour scale indicates the measured number of counts per ms. Every spectrum contains 28$\times$28 points, each averaged over 40 experimental realisations. The black curves show the eigenvalues of the undamped Hamiltonian. Lower row: Theory of Eqs.(\ref{eq:ar}) and (\ref{eq:flux}) after fitting to the data in the upper row. The fitting parameters are $C$ and the off-resonant amplitude.}
\label{fig:refl}
\end{figure}

We first consider the case of Fig.~\ref{fig:analogy}(a), where the cavity is probed directly. The upper row in Fig.~\ref{fig:refl} shows the number of photons detected per ms versus $\Delta_{ \rm a}$ and $\Delta_{ \rm c}$ for values of cooperativity ranging from $0.4$--$13.4$. The cooperativity was varied by changing the intra-cavity density of atoms, and $C$ was determined by fits to the spectra (shown in the bottom row), with the amplitude as the only other fit parameter. For these measurements, resonances of the coupled system are manifest as minima in the detected photocount rate as described by Eqs.(\ref{eq:ar}) and (\ref{eq:flux}). Note that the vertical axis covers a frequency range $\sim 1000 \times$ smaller than that of the vertical; the DIT resonances are much narrower than the bare cavity line width. Despite always operating in a regime where the cavity damping rate exceeds the Rabi frequency by an order of magnitude, the coherent effect of the atomic dipole is evident as soon as $C\sim1$. In particular it is seen that the reflection minima trace out the avoided crossing of the eigenvalues (\ref{eq:undampedeigens}) of the undamped Hamiltonian, satisfying the condition $\Delta_{ \rm a}=g^2/\Delta_{ \rm c}$. That the locations of these features are described independently of the decoherence rates $\gamma$ and $\kappa$ emphasizes the fact that the underlying coherent properties of the system have been revealed, despite the absence of Rabi oscillations.

\begin{figure}
\centering\includegraphics[width=0.95\textwidth]{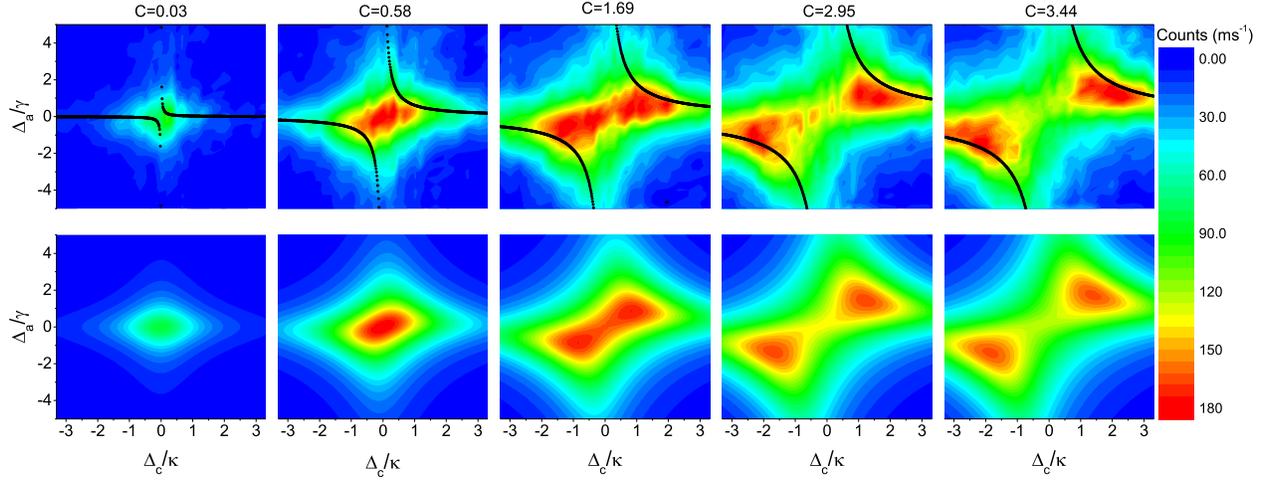}
\caption{{\bf Probing the system by driving the atoms.} Two dimensional scans of the cavity emission signal in the experimental configuration of Fig.~\ref{fig:analogy}(b) as a function of the detunings $\Delta_{ \rm c}$ and $\Delta_{ \rm a}$ for different values of the cooperativity $C$. Upper row: Experimental spectra. The colour scale reflects the measured number of counts per ms. Every spectrum contains 31$\times$31 points, each averaged over 40 experimental realisations. The black curves show the eigenvalues of the undamped Hamiltonian. Lower row: Theory of Eqs.(\ref{eq:af}) and (\ref{eq:flux}) after fitting to the data in the upper row with $j=0$. The fitting parameters are $C$ and the amplitude.} 
\label{fig:fluo}
\end{figure}

In the experimental configuration depicted in Fig. \ref{fig:analogy}(b) the coupled system is probed via scattering of incoming radiation by the atoms: the atomic dipole is driven by the probe beam from the side of the cavity and emits radiation at the same frequency into the resonator. According to Eq.~(\ref{eq:flux}) with $j=0$, the built up cavity field is transmitted through the coupling mirror, producing a steady-state flux of $2\kappa_T|a|^2$ photons per unit time, with $a$ given by Eq.~(\ref{eq:af}) and $\kappa_T\approx\kappa/2$ in our system. The detected photon count rate is shown in the upper row of Fig.~\ref{fig:fluo}. The values of $C$ here are lower than in reflection for comparable atomic densities because the side-driven atoms are more weakly coupled to the cavity mode \cite{Goldwin11}. For values of $C\lesssim 1$, the emission spectrum is localised around the origin of the $\Delta_{ \rm c}$--$\Delta_{ \rm a}$ plane. Comparing Eqs.(\ref{eq:ar}) and (\ref{eq:af}), we see that the driven atoms act as a secondary probe source exciting the cavity field with relatively small line width, $\gamma$:
\begin{eqnarray}
\eta &\to& \frac{-g\,\Omega/2}{\gamma-i\Delta_{ \rm a}} \quad.
\end{eqnarray}
As $C$ is increased, the atom-cavity coupling leads to a characteristic `butterfly' shape in the cavity emission spectrum. The central feature is split into two broad maxima along the scaled diagonal defined by $(\Delta_{ \rm a}/\gamma)=(\Delta_{ \rm c}/\kappa)\equiv\delta$, centred around $\delta=\pm\sqrt{C-1}$. As outlined above, in contrast with the reflection measurements, when the atoms are driven directly no DIT signature is expected. Therefore the resonance features no longer follow the eigenvalues of the undamped Hamiltonian (black curves). With the cooperativity and the amplitude of the signal as the only free parameters, we fit Eq.~(\ref{eq:af}) to our experimental data. The results are shown in the bottom row of Fig.~\ref{fig:fluo}.

\begin{figure}
\centering\includegraphics[width=0.48\textwidth]{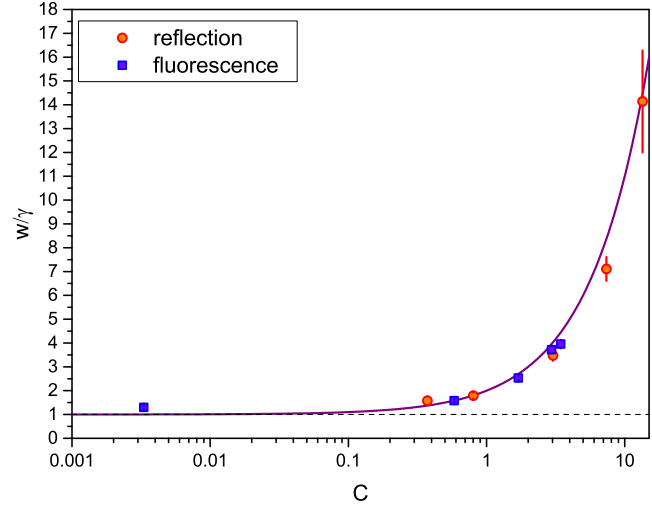}
\caption{{\bf Measured resonance width $w$ versus cooperativity $C$.} Data are extracted from the reflection and fluorescence spectra with $\Delta_{ \rm a}=\Delta_{ \rm c}$. The Lorentzian half-widths at half-maximum are normalized to the natural atomic line width $\gamma$, and error bars show the fit uncertainty. The dashed line is the asymptotic value $\gamma$, while the solid line shows the prediction $w=\gamma(1+C)$.}
\label{fig:width}
\end{figure}

Despite the fact that fluorescence measurements do not exhibit DIT, both probe configurations exhibit the same spectral width when the cavity resonance frequency is tuned to that of the atom in free space ($\Delta_{ \rm a}=\Delta_{ \rm c}$). This corresponds to nearly vertical traces through the plots in Figs.~\ref{fig:refl} and \ref{fig:fluo}. Along these traces, both types of spectra are well approximated by simple Lorentzians, and the spectral width $w$ is obtained as the half-width at half-maximum of a fit to such. The results are shown in Fig.~\ref{fig:width}, plotted against the cooperativity $C$ obtained from the full surface fits. The solid curve is the expectation $w=\gamma\,(1+C)$ with no free parameters. This is just the imaginary part of $\tilde{\omega}_-$, plotted in Fig.~\ref{fig:eigenfreqs}(b) as the lower (blue dashed) branch. For the reflection measurements we understand this peak as the dressing-induced transparency window described above, having a deeply sub-natural line width ($\ll\kappa$) dependent on the coherent coupling rate $g$. For fluorescence, the atoms scatter coherently into the cavity mode at a rate equal to $\gamma\,C$ through the Purcell effect. Following the abrupt shut-off of a resonant probe laser, one would observe that the intracavity power would decay exponentially at a rate $2w$ for times longer than $1/\kappa$. The fact that this decay is much slower than the bare cavity ring-down rate shows that the relatively long coherence time of the atoms  effectively inhibits the decay of the coupled system.

\section{Discussion}

We have presented an experimental study of coherence within an overdamped quantum system consisting of two-level atoms coupled to an optical microcavity. Despite the absence of Rabi oscillations in our system, the cavity reflection signal reveals the avoided crossing associated with the eigenvalues of the undamped Hamiltonian. We note that the response of the intracavity field to a probe beam is formally identical to that of a three level atom in free space exhibiting electromagnetically induced transparency. This provides insight into the persistent role of coherence in the so-called bad cavity regime of cavity QED. We prefer to call this the fast cavity regime. We have also shown that, although this DIT effect is intrinsically absent when the atoms are driven directly from the side of the cavity, both experimental configurations produce a split spectrum when the cooperativity becomes large enough.

The formal mapping onto EIT also suggests a range of applications for fast cavity QED with cold atoms on hybrid atomic-photonic chips. These include dipole-induced switching \cite{Waks06,Faraon08} and optical pulse lag/lead circuits \cite{Shimizu02,Rao04,Auffeves07}. In the absence of atoms, a resonant cavity provides a group delay of $1/\kappa$ in transmission; as the coupling $g$ is increased, the delay is reduced and becomes a superluminal advance approaching $1/\gamma$ for $g^2\gg\gamma^2$ (see Methods). This can be understood by considering the phase evolution of light propagating through the atomic medium in the absence of the cavity. For an absorption resonance, the shape of the imaginary part of the optical susceptibility is inverted relative to a transmission resonance. Then the Kramers-Kronig relations imply that the dispersion must be anomalous in order to preserve causality. For our experiment, operating deep in the Purcell regime, this anomalous phase evolution leads to a narrow region of negative group delay within the DIT window. In some ways this is reminiscent of the anti-resonance observed in \cite{Sames14}. However there are two essential differences between that work and ours. First, observation of the anti-resonance relies on strong coupling to scatter enough light into the cavity mode to interfere destructively with the incident probe. In contrast, it is a Fano-type interference between two decay channels coupled to a single reservoir which underlies the DIT resonance \cite{Pen14,Ani11,Fleisch05}. Second, the characteristics of the anti-resonance depend solely on the atomic properties, while we have shown that the DIT resonance mirrors the avoided crossing of the normal modes. This makes DIT attractive for the detection of coherence in systems where strong coupling is either not feasible or undesired.

\section{Methods}

\subsection{Experiment}
The apparatus has been described previously \cite{Tru07,Goldwin11}, so will only be summarised here. A magneto-optical reflection trap (reflection-MOT), formed above a gold mirror, collects a few million $\mathrm{^{87}Rb}$ atoms in a cloud of approximate size $\mathrm{1 \times 1\times 2\ mm}$. The cloud is released, cooled further in sub-Doppler optical molasses, then pushed with a pulsed resonant laser beam through a 1\,mm hole in the mirror and into the optical microcavity, which lies approximately 5\,mm below. 

The microcavity itself is a plano-concave Fabry-P\'erot resonator, built as described in \cite{Tru05}. The plane mirror is a multi-layer dielectric stack that we have glued onto the tip of a cleaved single-mode optical fibre. This mirror serves as the input-output coupling mirror. The second mirror is one of an array of approximately spherical micro-mirrors with $185~\mu$m radius of curvature on an etched and dielectric-coated silicon wafer. The cavity length is tuned around $127~\mu$m using a shear piezo on which the fibre mirror is mounted, and is stabilised with a side-of-fringe technique using light from a single-frequency Ti:sapphire laser (Coherent, MBR-110) operating at $800\,$nm. The wavelengths of both locking and probing lasers are monitored with a wavemeter (HighFinesse, WS6). The cavity finesse varies slightly throughout the measurements presented here due to a slow drift in the mirror alignment. The finesse (line width $\kappa/(2\pi)$) is 260 ($2.2$~GHz) in Fig.~\ref{fig:refl}, and 180 ($3.2$~GHz) in Fig.~\ref{fig:fluo}. The fringe contrast, defined by $1-$(min power)/(max power) throughout a cavity scan, is above 95\% for all of the measurements.

Near-resonant signal light from the cavity is separated from the locking light and incoming probe light using a 90/10 beam splitter and a series of dichroic filters, and detected on a single-photon counting module (Perkin-Elmer, SPCM-AQR-14). The detected counts are corrected for dead time at high flux rates with a polynomial formula as specified by the manufacturer.

\subsection{Group delay}

The group delay $T_{ \rm g}$ corresponding to transmission through a cavity can be obtained from the cavity transfer function via
\begin{eqnarray}
T_{ \rm g} &=& \frac{\partial}{\partial\omega_{\rm p}}\arg(a/\eta)\quad. %\left(\frac{a}{\eta}\right)\quad.
\end{eqnarray}
Taking $a$ in Eq.~(\ref{eq:ar}) with $\Delta_{ \rm a}=\Delta_{ \rm c}$, and Taylor expanding around resonance, one obtains
\begin{eqnarray}
T_{ \rm g} &=& \frac{1}{\kappa}\,\frac{1-g^2/\gamma^2}{1+g^2/(\kappa\gamma)} \quad,
\end{eqnarray}
giving the limits
\begin{eqnarray}
T_{ \rm g} &\to& \left\{ \begin{array}{rl}
1/\kappa, & \quad g\ll\gamma \\
0, & \quad g=\gamma \\
-1/\gamma, & \quad g\gg\gamma \end{array}\right. \quad.
\end{eqnarray}
The sign of the group delay for the dressed cavity in DIT is opposite to that in conventional EIT, where $T_{ \rm g}$ becomes large and positive as the transparency is increased. In the strong DIT limit (with $\kappa\gg g\gg\gamma$), one recovers the relatively large group advance, equal to $1/\gamma$, that an atomic medium with equivalent optical depth would produce in free space. As in free space, there is an intrinsic compromise between group dispersion and transmission in this limit of DIT, due to the suppression of the latter when $C$ becomes large.

\subsection{Data Availability}

The data presented here are available from the research data management system of the University of Birmingham, accessible online at http://epapers.bham.ac.uk/2213/.

% Create the reference section using BibTeX:
%\bibliography{basename of .bib file}

% If you have acknowledgments, this puts in the proper section head.
\section{Acknowledgements}
This work was funded by the EU (FP-7 grant 247687), by the UK EPSRC, and by the Royal Society. M.S.~acknowledges financial support from the Austrian Science Fund (DK CoQuS project No.~W 1210-N16) and M.M.~from the Stiftung der Deutschen Wirtschaft (SDW). We thank Jon Dyne and Val Gerulius for technical expertise, and Michael Trupke, Joanna Kenner, and Martin Everett for early contributions to the apparatus. We gratefully acknowledge stimulating discussions with Shuang Zhang and the late Wolfgang Lange.

\section{Contributions}
The experiments were conceived and supervised by J.G.~and E.A.H. Data were obtained by Y.-H.L.~with the help of G.B., M.S., and M.M., and analysed by Y.-H.L.~and G.B. All authors contributed to writing and editing the manuscript.

\section{Competing financial interests}
The authors declare no competing financial interests.

\newpage

\newpage

\newpage

\newpage

\newpage

\end{document}